\documentstyle[twocolumn,amssymb,aps,psfig]{revtex}
\tighten
\begin{document}
\draft

\title{\bf Bound-states in dimerized and frustrated Heisenberg Chains}

\author{G.~Bouzerar, S.~Sil}

\address{ 
Institut f\"ur Theoretische Physik, Universit\"at zu K\"oln,\\  
Z\"ulpicher Str. 77, D--50937 K\"oln, Germany \\
 }
\address{~
%\begin{abstract}
\parbox{14cm}{\rm
\medskip
Using the Bond Operator Technique (BOT), we have studied the low 
energy excitation
spectrum of a frustrated dimerized antiferromagnetic Heisenberg chain.
In particular, we have compared our analytical results with previous Exact Diagonalizations (ED) data. 
Qualitatively, the BOT results are in good agreement with the
ED data. And even a very good quantitative agreement is obtained in some parameter region. It is clearly shown that there is only one elementary excitation branch (lowest triplet branch) and that the two other 
well defined excitations which appear below the continuum, one singlet and
one triplet, are bound states of two elementary triplets.
\\ \vskip0.05cm \medskip PACS numbers: 71.27.+a, 75.40.Mg, 75.90.+w
}}
%\end{abstract} 
\maketitle

\narrowtext

\section{INTRODUCTION}
The magnetic properties of low dimensional quantum spin systems 
have attracted a considerable attention of theorists over the last decades. 
The interest for these systems was boosted by the recent discovery of non 
organic spin-Peierls materials like $CuGeO_3$ \cite{Hase} and 
$NaV_{2}O_{5}$\cite{Isobe,Weiden} and spin ladder compounds 
like $SrCu_2O_3$, $Sr_2Cu_3 O_5$ \cite{Hiroi,Dagotto}. 
It turns out that some of the properties of these 
materials can be described by the frustrated dimerized Heisenberg model. 
In the one dimensional case, the Hamiltonian of this model reads,

\begin{equation}
H=J\sum_{i}\left([1+\delta(-1)^i]{\bf S}_i\cdot{\bf S}_{i+1} + 
\alpha{\bf S}_i \cdot {\bf S}_{i+2}\right)
\label{hamilt}
\end{equation}
where $i$ denotes the sites of a chain with length $L$ and 
${\bf S}_i$ are $S= 1/2 $ spin operators. $J>0$ is the nearest--neighbor 
exchange coupling, $\alpha$ the frustration parameter from next--nearest 
neighbor coupling and $\delta$ is the dimerization parameter.

Depending on the coupling parameters, this model exhibits various phases including spin-liquid phase (gapless phase),
dimerized phase, Neel ordered phase and incommensurate phase.
The ground-state properties of this model were extensively studied using 
various numerical and analytical tools, like Exact Diagonalizations method, 
Quantum Monte Carlo, DMRG (density matrix renormalization group), Bethe ansatz, 
bond operator technique, etc..
Another interesting feature of the frustrated dimerized Heisenberg model is the
possible existence of well defined excited states below the continuum (bound-states?) \cite{Uhrig,Bouzerar1,Bouzerar2,Barnes}.
Recently, Bouzerar et al. have provided a detailed numerical analysis (Exact diagonalizations method) of the excitation spectrum in this model 
\cite{Bouzerar2}. It was shown that:
(i)The Singlet-Triplet excitation gap ratio is a universal 
function which depends on the frustration parameter only and 
(ii)A well defined second triplet branch appears below the continuum
in the vicinity of momentum $q=0$ (resp. $\pi$) only if the strength of the frustration is large enough.
A useful approach to describe the dimerized phase is the Bond-Operator representation of spins introduced by Chubukov \cite{Chubukov} and Sachdev and Bhatt
\cite{Sachdev}. Recently a number of works have been carried out with the 
help of bond-operator method. Employing the bond-operator methodology 
Brenig \cite{Brenig} studied the effect of interchain coupling in spin-Peierls
systems like $CuGeO_3$.
This method was also successfully used by Sushkov and Kotov to spin 
ladder model in the strong interchain coupling limit $J_{\perp} >> J{//}$, 
to show the existence of a singlet and triplet bound states below the 
continuum \cite{Kotov2}.
In this paper we will study the excited states of the frustrated dimerized 
Heisenberg model using the Bond-Operator Technique (BOT) and compare the results of our calculation with the Exact Diagonalizations data (ED) 
\cite{Bouzerar2}.
We will show that the agreement between ED and BOT is very good at least 
qualitatively, and even quantitatively in the vicinity of momentum $q=\pi/2$.
We will also show that the lowest triplet branch is the only elementary 
excitation and that the singlet and the triplet which appear below 
the continuum are bound-states of two triplets.

\section{Bond Operator technique}

In this method the Hilbert space of the spin degrees of freedom is 
represented  in terms of singlet and three triplet states which is created 
out of vacuum by one singlet ($ s^\dagger$) and three triplets ($ t^\dagger$) 
creation operators

\begin{eqnarray}
| s_i \rangle &=& s_i^\dagger |0 \rangle =  \frac{1}{\sqrt{2}}
 (| \uparrow_{2i} \downarrow_{2i+1} \rangle - | \downarrow_{2i} 
\uparrow_{2i+1} \rangle ),
 \nonumber \\  
| t_{x i} \rangle &=& t_{ x i}^\dagger |0 \rangle = - \frac{1}{\sqrt{2}}
 (| \uparrow_{2i} \uparrow_{2i+1} \rangle - | \downarrow_{2i} 
\downarrow_{2i+1} \rangle ),
 \nonumber \\  
| t_{y i} \rangle &=& t_{ y i}^\dagger |0 \rangle =  \frac{i}{\sqrt{2}}
 (| \uparrow_{2i} \uparrow_{2i+1} \rangle + | \downarrow_{2i} 
\downarrow_{2i+1} \rangle ),
 \nonumber \\  
| t_{z i} \rangle &=& t_{z i}^\dagger |0 \rangle =  \frac{1}{\sqrt{2}}
 (| \uparrow_{2i} \downarrow_{2i+1} \rangle + | \downarrow_{2i} \uparrow_{2i+1}
 \rangle ).
\end{eqnarray}

 A representation of spins in terms of the singlet and triplet operators
 is given by 
\begin{eqnarray}
S_{\alpha 2i} &=& \frac{1}{2} (s_i^\dagger t_{\alpha i} + t_{\alpha i}^\dagger
s_i - i \epsilon _{\alpha \beta \gamma} t_{\beta i}^\dagger t_{\gamma i}), 
\nonumber  \\
S_{\alpha 2i+1} &=& \frac{1}{2} (-s_i^\dagger t_{\alpha i} - 
t_{\alpha i}^\dagger
s_i - i \epsilon _{\alpha \beta \gamma} t_{\beta i}^\dagger t_{\gamma i}), 
\end{eqnarray}

where $\alpha$, $\beta$ and $\gamma$ = $x,y,z$ and $\epsilon_{\alpha
 \beta \gamma} $ is the Levicivita
symbol representing the totally antisymmetric tensor. Henceforth it is assumed 
that all repeated indices over $\alpha, \beta$ and $\gamma$ are summed over.
The four operators ($s, t_x, t_y$ and $t_z $) satisfy the usual bosonic 
commutation relations. In order to ensure that the physical states are either 
singlet or triplets one has to impose the constraint
\begin{equation}
s_i^\dagger s_i + t_{\alpha i}^\dagger t_{\alpha i} = 1.
\label{constraint}
\end{equation} 
It is a difficult task to fulfill this constraint exactly (in each bond) and
in most of the works this constraint is satisfied only in average.
To overcome this difficulty, an infinite on site repulsion has been introduced
by Kotov et al. \cite{Kotov1}.
In Bond-operator representation the Hamiltonian reads,

\begin{eqnarray}
 H &=& H_0 + H_1 +H_2
\label{eqhamilt}
\end{eqnarray}
\begin{eqnarray}
H_0 &=& - \frac{3}{4} J (1+\delta) \frac{L}{2} + J (1+\delta) \sum_{i=1,\alpha}^
{\frac{L}{2}} t_{\alpha ,i}^{\dagger} t_{\alpha ,i} \nonumber \\
&-&K_{0} \sum_{i=1}^{\frac{L}{2}} (t_{\alpha i}^{\dagger}t_{\alpha i+1}s_{i}^
{\dagger} s_{i+1} + t_{\alpha i}^{\dagger}t_{\alpha i+1}^\dagger s_is_{i+1}
+h.c)  \nonumber \\
H_1 &=& K_{1} \sum_{i=1, \alpha \neq \beta} ^
{\frac{L}{2}}( t_{\alpha i}^\dagger t_{\beta i+1}^\dagger t_{\beta i}t_{\alpha
i+1} - t_{\alpha i}^\dagger t_{\alpha i+1}^\dagger t_{\beta i} t_{\beta i+1})
\nonumber \\
H_2 &=& K_{2} \sum_{i=1 \alpha, \beta,\gamma}^{\frac{L}{2}}
(i \epsilon_{\alpha \beta \gamma} [t_{\alpha i}^\dagger t_{\beta i+1}^\dagger
t_{\gamma i+1}s_{i} \nonumber \\
&-& t_{\alpha i+1}^\dagger t_{\beta i}^\dagger
t_{\gamma i}s_{i+1}] +h.c) \nonumber 
\end{eqnarray}

with $K_{0}=\frac{J}{4}(1-\delta -2 \alpha )$,$K_{1}=\frac{J}{4}(1-\delta 
+2 \alpha )$ and $K_{2}=\frac{J}{4}(1-\delta)$.

This method gives the exact ground-state in the limit when the ground state
wave function consists of a product of local independent dimers which
 is especially realized on the 'disordered line'' $2 \alpha + \delta =1$.
On this line the Ground-state energy per site is $E_{G}/L= -\frac{3}{8} 
(1 + \delta)$.
It should be noted that in the BOT the elementary excitation is a local triplet (nearest neighbor sites).

In order to perform the calculation of the triplet dispersion:
(i) Initially we neglect completely the local constraint which is 
equivalent to set $s_i=1$.
(ii) We restrict ourself to the quadratic terms only. As it was previously 
shown that the effect of the higher order terms $H_{1}$ and $H_{2}$ are small 
\cite{Gopalan,Sachdev}.

After using the well known Bogoliubov transformation $t_{\alpha k}=
u_{k}a_{\alpha k}+ v_{k}a_{\alpha -k}^{\dagger}$, the hamiltonian reduces to,

\begin{eqnarray}
 H &=& -\frac{9}{8}J (1+\delta)L + \sum_{\alpha,k=\frac{-\pi}{2}}^{\frac{\pi}{2}}
\omega_{k}(a_{\alpha k}^{\dagger}a_{\alpha k}+\frac{1}{2})
\end{eqnarray}

where
\begin{eqnarray}
\omega_{k}&=&\sqrt{A_{k}^{2}-B_{k}^{2}} \\
u_{k}&=&\sqrt{\frac{1}{2} + \frac{A_{k}}{2\omega_{k}}} \nonumber \\
v_{k}&=&\sqrt{-\frac{1}{2} + \frac{A_{k}}{2\omega_{k}}}\nonumber 
\end{eqnarray}

with $A_{k}=J(1+\delta)-\frac{J}{2}(1-\delta-2\alpha)cos(2k)$ and 
$B_{k}=-\frac{J}{2}(1-\delta-2\alpha)cos(2k)$.

Let us now take into account the effect of the local constraint 
eq. (\ref{constraint}). Following the ref. \cite{Kotov1} we include in the
previous Hamiltonian an infinite on site repulsion $H_{U}$,

\begin{eqnarray}
H_{U}=U \sum_{i,\alpha \beta}t^{\dagger}_{\alpha i}t^{\dagger}_{\beta i}
t_{\beta i}t_{\alpha i},~~~~ U \rightarrow \infty
\end{eqnarray}

Concerning the calculation of the self-energy correction to the dispersion
we follow the diagrammatic approach developed in ref. \cite{Kotov1}.
Let us just summarize the main steps of the calculation.
At first the vertex scattering amplitude $\Gamma(k,\omega)$ is evaluated 
by solving 
the Bethe-Salpeter equation shown in fig.\ref{fig1} a.

\begin{eqnarray}    
&&\Gamma(k,\omega) = [ i\int dq d \omega' G(q,\omega') G(k-q, \omega 
- \omega')]^{-1} \nonumber \\
&=& -[ \frac{1}{N} \sum _q \frac{u_q^2 u_{k-q}^2}{\omega -\omega_q - 
\omega_{k-q}} + \{ u \rightarrow v , \omega \rightarrow -\omega \}] ^{-1}
\end{eqnarray}

where $G(k, \omega)$ is the normal Greens function of $H_0$.
We neglect the higher order terms of $v_q$ since  
$\sum _{q}v_q^2$ is proportional to the number of triplets in the system and it
is very small in the dimerized phase. 

Then taking the scattering vertex as $\Gamma(k, \omega)$, the corresponding 
self energy is obtained from the diagram shown in fig 1 b,
\begin{equation} 
\Sigma (k, \omega) = \frac{8}{N} \sum_{q= - \frac{\pi}{2}}^{\frac{\pi}{2}}
v_q^2 \Gamma (k+q, \omega - \omega_q) .
\end{equation}

Now we calculate the renormalized spectrum 
\begin{equation}
\Omega_k = Z_k \sqrt{((A_k + \Sigma (k,0))^2 - B_k^2)}
\label{eqomega}
\end{equation}
 from the poles of the modified Greens function. Here  
the quasiparticle weight 
\begin{equation}
Z_k = ( 1- \frac{\partial \Sigma}{\partial \omega})^{-1}
\end{equation}

\section{Singlet-Triplet gap and Dispersion}

In this section we will compare the BOT results (with and without the 
constraint) of the Singlet-Triplet gap and Dispersion to
the ED data of ref.\cite{Bouzerar2}.
We observe in fig. \ref{fig1}, that the gap $\Delta_{BOT}$ is always larger than the exact result and even gets worse when the constraint is included.
For a given $\delta$, the difference between the gaps with and without 
constraint reduces when it approaches the disordered line.
On the disordered line, it can be seen easily from equation (\ref{eqomega}), that
the dispersion is unchanged when the constraint is included, since
$\Sigma(k,\omega) \propto (2\alpha+\delta -1)$.
However, the agreement between BOT and ED gets better in the limit of 
large $\delta$. For instance at $\alpha=0$, we get 
$\frac{\Delta_{ED}}{\Delta_{BOT}}= 0.67$ for $\delta=0.2$, but  
this ratio is 0.87 when $\delta=0.4$.  This ratio increases with $\alpha$.
Furthermore, the curve including the constraint is parallel
to the ED data curve, this means that the effect of frustration is 
taken into account more properly when the constraint is included.
However, in the limit of vanishing $\delta$ the effect of frustration
is not taken properly in the BOT approach \cite{Brenig}. 
For instance at the Majumdar-Ghosh \cite{Majumdar}
point $(\alpha=0.5, \delta=0.)$, $\frac{\Delta_{ED}}{\Delta_{BOT}} \approx 0.25$.

Let us now discuss the dispersion of the triplet excitation ($\omega(q)$).
In order to analyze the effect of the frustration on $\omega(q)$ we have
fixed the dimerization parameter $\delta=0.2$.
In fig. \ref{fig3} a (resp.\ref{fig3} b) we have plotted the dispersion for 
$\alpha=0$ (resp. $\alpha=0.3$).
In fig.\ref{fig3} a and fig.\ref{fig3} b  we observe that the agreement 
between the ED data and BOT with constraint are excellent in the vicinity 
of $q=\pi/2$. This agreement will get better with increasing $\delta$.
Note that on the disordered line the BOT data coincides exactly with the ED data. Indeed, on this line and at $q=\pi/2$, the lowest triplet excitation is
$|T(\pi/2) > = \sum_{l} e^{i\frac{\pi}{2}l} S^{\dagger}_{l} |GS> $, where the 
ground-state wave function $|GS>$ consists of product of independent dimers 
\cite{Shastry}.
Note that even far from the disordered line, this state remains a good approximation of the exact lowest triplet state in the vicinity of $\pi/2$.
It is important to emphasize, that away from $q=\pi/2$, the disagreement between BOT and ED data is mainly due to the fact that the 
{\it{triplet excitation are not local objects}}, but in the BOT scheme by construction the triplet excitation are local.
We observe in fig.\ref{fig3} a (resp.\ref{fig3} b) that when
approaching the disordered line the width of the dispersion strongly decreases,
and the agreement with the ED data gets worse.
On the disordered line, the BOT triplet excitation is dispersionless and $\omega(q)=J(1+\delta)$.
However, as discussed previously, when $\delta$ is increased the agreement 
between BOT and ED gets better
and better and coincide for the special case $\delta=1$ and $\alpha=0$.

\section{Two magnons bound-state.}

The quartic term $H_{1}$ in the Hamiltonian eq. \ref{eqhamilt},
consist of triplet-triplet interaction. This term can lead to an attractive
interaction between triplets to form bound-states.
This bound-states can be singlet, triplet or quintuplet.
We will focus only on the possibility of a singlet or triplet bound-states
which were observed below the continuum in the ED calculations \cite{Bouzerar2}.
In the previous section we have seen that the effect of the constraint 
does not significantly modify the triplet dispersion, thus to simplify the calculations we will neglect the constraint.

\subsection{Singlet bound-state.}

To perform the calculation of the singlet bound-state , let us consider the 
most general singlet wave function 

\begin{equation} 
| \Psi^{S}(Q)>=\sum_{k,\alpha} \Phi^{S}(k) a_{\alpha Q/2-k}^{\dagger}a_{\alpha Q/2+k}^{\dagger} |0 >
\end{equation}
of total momentum Q, constructed from two triplet elementary excitations, where 
$\Phi^{S}(k)$ is determined from the Schr$\ddot{o}$dinger equation,

\begin{equation} 
H | \Psi^{S}(Q)> =E^{S}(Q)| \Psi^{S}(Q) > 
\end{equation}
with singlet bound state energy $E^{S}(Q)$. 
This leads immediately to the integral equation of the form,
\begin{eqnarray} 
(E^{S}(Q)-\omega_{Q/2-q} &-& \omega_{Q/2+q}) \Phi^{S}(q) \nonumber \\
&=& -4 K_{1}g(q) \int_{-\frac{\pi}{2}}^{\frac{\pi}{2}}
 \frac{dp}{\pi} g(p)\Phi^{S}(p)
\label{eqbs0}
\end{eqnarray}
where $g(p)=cos(2p)u_{ Q/2-p}u_{ Q/2+p}$.

In the vicinity of the disordered line ($|2\alpha+\delta-1| << 1$) this integral can be solved analytically.
In this region of parameter $u_k \approx 1$ and $\omega_{k}/J \approx (1+\delta)-\frac{1}{2} (1-\delta -2\alpha)cos(2k)$.
The corresponding singlet energy for the  momentum Q is,

\begin{eqnarray} 
& & E^{S}(Q)/J = 2(1+\delta)  \nonumber \\
& - & \frac{1}{2}\sqrt{ (1-\delta+2\alpha)^{2}+4(1-\delta-2\alpha)^{2} cos^{2}(Q)} .
\end{eqnarray}

For any momentum Q the energy of the singlet state is always smaller than the energy of the lower edge of the continuum defined as $E^{C}(Q)=min_{q}(\omega_{Q/2-q}+\omega_{Q/2+q})=2(1+\delta)- |(1-\delta-2\alpha)cos(Q)|$.
Thus, a well defined singlet bound-state exists for any momentum Q.
Far from the disordered line the equation (\ref{eqbs0}) should be solved 
numerically.

\subsection{Triplet bound-state.}

Let us now perform analogous calculations in the triplet sector.
The most general triplet wave function can be written,

\begin{equation} 
| \Psi^{T}_{\alpha}(Q)>=\sum_{k,\beta \gamma} \Phi^{T}_{\alpha}(k)
 \epsilon _{\alpha \beta \gamma}a_{\beta Q/2-k}^{\dagger}a_{\gamma Q/2+k}^
{\dagger} |0 >.
\end{equation}

The corresponding Schr$\ddot{o}$dinger equation is,

\begin{eqnarray} 
(E^{T}(Q)-\omega_{Q/2-q} & - & \omega_{Q/2+q}) \Phi^{T}_{\alpha}(q)= 
\nonumber \\
& -& 2 K_{1}h(q) \int_{-\frac{\pi}{2}}^{\frac{\pi}{2}}
 \frac{dp}{\pi} h(p)\Phi^{T}_{\alpha}(p)
\label{eqbs1}
\end{eqnarray}
where $h(p)=sin(2p)u_{ Q/2-p}u_{ Q/2+p}$.

As we did it previously, we solve this equation analytically in the vicinity 
of the disordered line.
We get,
\begin{equation} 
E^{T}(Q)/J=2(1+\delta)-K_{1}[1+(\frac{2K_{0}}{K_{1}})^{2}cos^{2}(Q)]
\end{equation}

We find that the Triplet is below the continuum (i.e. $E^{T}(Q) < E^{C}(Q)$)
when $Q > Q_{c}$, where

\begin{equation} 
 Q_{c}=cos^{-1}[\frac{1-\delta+2\alpha}{2(1-\delta-2\alpha)}]
\label{eqqc}
\end{equation}

\subsection{Discussions.}

In this subsection we will compare the BOT results to the ED data.
Let us first discuss the dependence of the singlet-triplet ratio 
$R=\frac{\Delta^{S}}{\Delta^{T}}$, where $\Delta^{S}$ is the singlet-singlet
gap and $\Delta^{T}$ the singlet-triplet gap ($\Delta^{T}=\Delta$).
It was shown in ref.\cite{Bouzerar2} that, in presence of dimerization 
($\delta \neq 0$) this ratio is a universal function
which depends on the frustration parameter only (in the commensurate region).
In fig. \ref{fig4} we have plotted R for different parameter $\delta$ as a function of $\alpha$. The agreement with the ED data is surprisingly good.
We observe that $R^{BOT}$ has a small dependence on $\delta$. However, 
when $\delta$ is increasing $R^{BOT} \rightarrow R^{ED}$. Especially for 
$\alpha=0$ we observe that $R(\alpha =0) \rightarrow 2 $.
The deviation for large $\delta$ in the ED data were attributed to the 
crossing of the disordered line.

It was shown in ref.\cite{Bouzerar2}, that a well defined second triplet branch
split from the continuum in the vicinity of $q=0$ (resp. $q=\pi$) if the 
strength of the frustration is large enough, $\alpha > \alpha^{*}(\delta)$.
In particular it was shown that $\alpha^{*}(\delta) \approx (1-\delta)/3$.
In fig.\ref{fig5} we have plotted $\alpha^{*}$ as a function of $\delta$
calculated exactly and within the BOT method.
We find that the qualitative agreement with the ED data is very good. For large $\delta$ we have found that $\alpha_{BOT}^{*}(\delta) \approx 1/2 \alpha_{ED}^{*}(\delta)$. We believe that this discrepancy on the slope is due
to the fact that the width of the dispersion of the elementary triplet is
underestimated in the BOT approach.

We observe that in the unfrustrated case ($\alpha=0$) increasing $\delta$ reduces the region where the second triplet appears below the continuum.
In other words, $Q_c(\delta)$, the momentum where the triplet split from 
the continuum, increases with $\delta$.
The variation of $Q_{c}$ with $\delta$ is plotted in fig.\ref{fig6} and it is
in agreement with eq.(\ref{eqqc}), $Q_{c} \rightarrow \pi/3$ when $\delta 
\rightarrow 1$.
Thus, in the unfrustrated case, the effect of $\delta$ {\it{reduces}} 
the effective interaction between two elementary triplets excitation.
In order to visualize the effect of the frustration, we have plotted 
in fig.\ref{fig7}, the dispersion of the lowest excitations, ie. the 
elementary triplet excitation and the two bound-states for a fixed value of 
the dimerization parameter. The figure shows,
in agreement with the previous work \cite{Bouzerar2}, that the effect of 
the frustration increases the region where the second triplet is a well 
defined excitation. These figures are very similar to fig.5 of ref.\cite{Bouzerar1}.

\section{Conclusions.}

As a conclusion, using the Bond operator method we have shown that in the frustrated dimerized Heisenberg model, there is only one elementary excitation branch (lowest triplet branch).
Depending on the parameters ($\delta,\alpha$), two bound states of these 
 elementary excitations, one singlet and one triplet, can appear below the continuum. Even in absence of frustration the singlet bound-state is a well
defined excitation for any momentum. However, in good agreement with previous work, the region where the second triplet is well defined depends on the frustration parameter. The triplet bound-state is observable in the vicinity of $q=0$ (resp. $q=\pi$) when the frustration strength is large enough.
Furthermore, in the unfrustrated case, the region where the second triplet 
is split from the continuum, reduces when
increasing the dimerization parameter $\delta$.

\centerline{{\bf Acknowledgments}}
We gratefully acknowledge helpful discussions with W. Brenig, and E. M\"uller-Hartmann.  This research was performed within the program of the Sonderforschungsbereich 341 supported by the Deutsche Forschungsgemeinschaft.

%
%  Fig. 1
%

\begin{figure}
\caption[]{Diagrams in the ladder approximation for the vertex scattering 
amplitude (a), and the self-energy (b)
}
\label{fig1}
\end{figure} 

%
%  Fig. 2
%

\begin{figure}
\caption[]{Gap $\Delta$ as a function of $\alpha$ for $\delta=0.2$ and 0.4
.The full symbols are Exact Diagonalizations data from ref. \cite{Bouzerar2}
(extrapolated in the thermodynamic limit).
The continuous line are BOT calculations without constraint, and the dotted line
with constraint.
}
\label{fig2}
\end{figure} 

%
%  Fig. 3
%

\begin{figure}
\caption[]{Dispersion of the lowest triplet branch for a fixed $\delta=0.2$ and $\alpha=0$ (a) and $\alpha=0.3$ (b).
The symbols are ED data (for a system size $L=20$).
The dashed line are BOT calculations without constraint, and the continuous line with constraint.
}
\label{fig3}
\end{figure} 

%
%  Fig. 4
%

\begin{figure}
\caption[]{Singlet-Triplet gaps ratio.
The symbols are ED data from ref. \cite{Bouzerar2}.
The lines are BOT data calculated for different values of $\delta$.
}
\label{fig4}
\end{figure} 

%
%  Fig. 5
%

\begin{figure}
\caption[]{$\alpha^{*}$ as a function of $\delta$.
Symbols are ED data from ref. \cite{Bouzerar2}, and the continuous line corresponds to BOT results.
}
\label{fig5}
\end{figure}

%
%  Fig. 6
%

\begin{figure}
\caption[]{$Q_{c}$ as a function of $\delta$ in the unfrustrated case.
}
\label{fig6}
\end{figure} 

%
%  Fig. 7
%

\begin{figure}
\caption[]{Dispersion of the triplet elementary excitation, singlet bound-state
and triplet-bound-state for a fixed $\delta=0.2$ and $\alpha=0.1$ (a),0.17 (b) and 0.25 (c).
The continuum corresponds to the shaded region.
}
\label{fig7}
\end{figure}

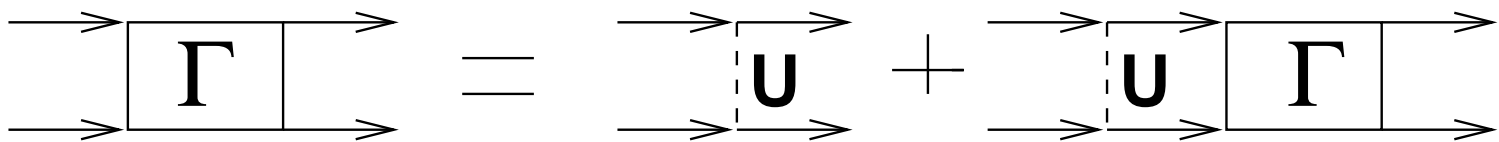
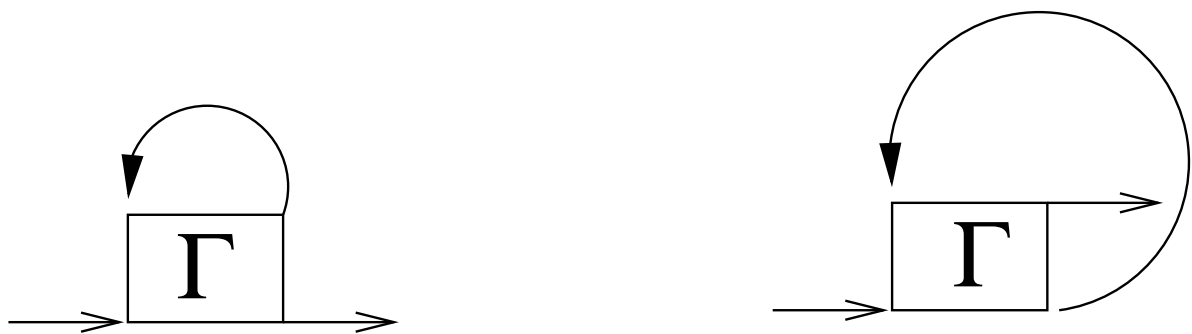
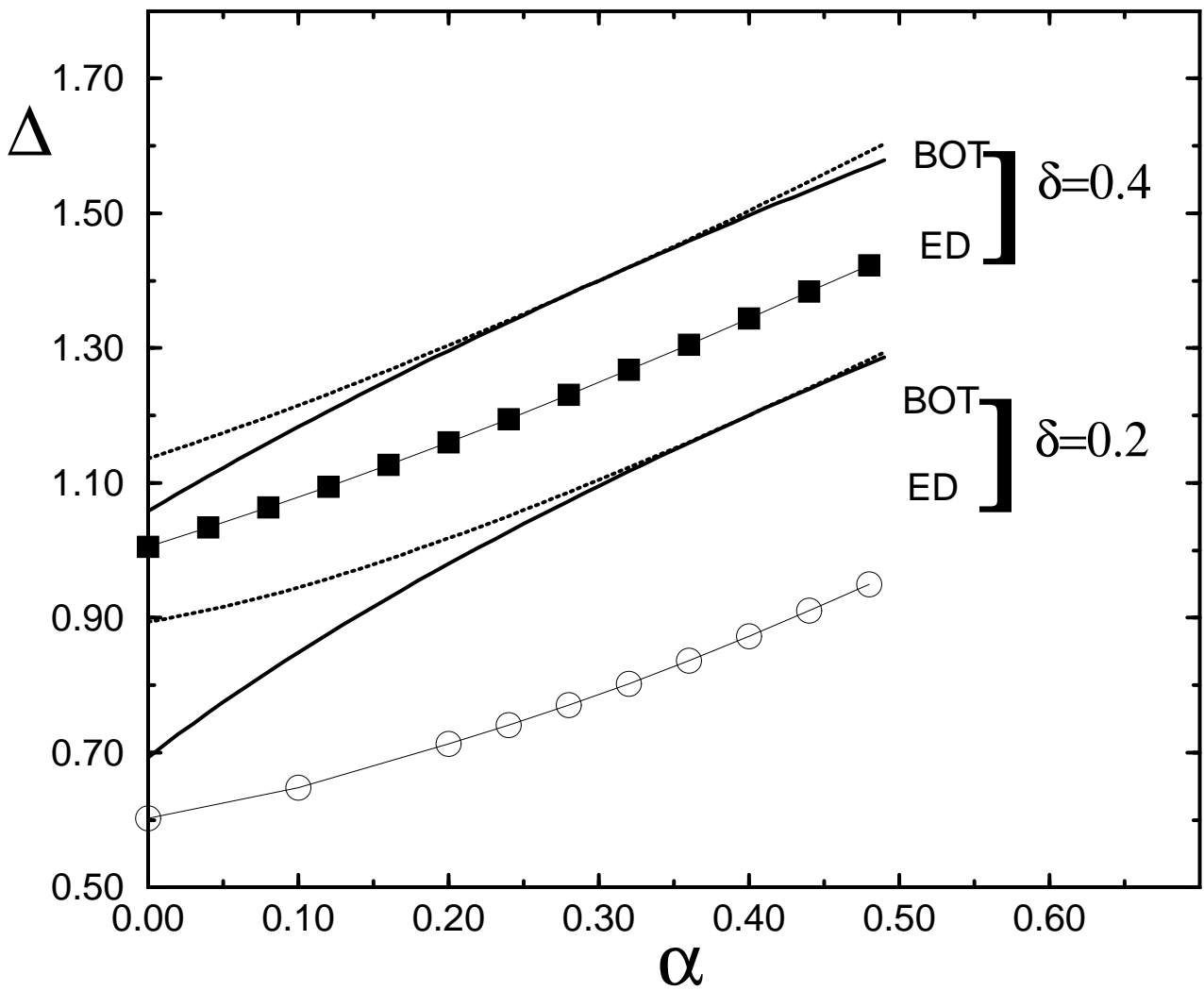
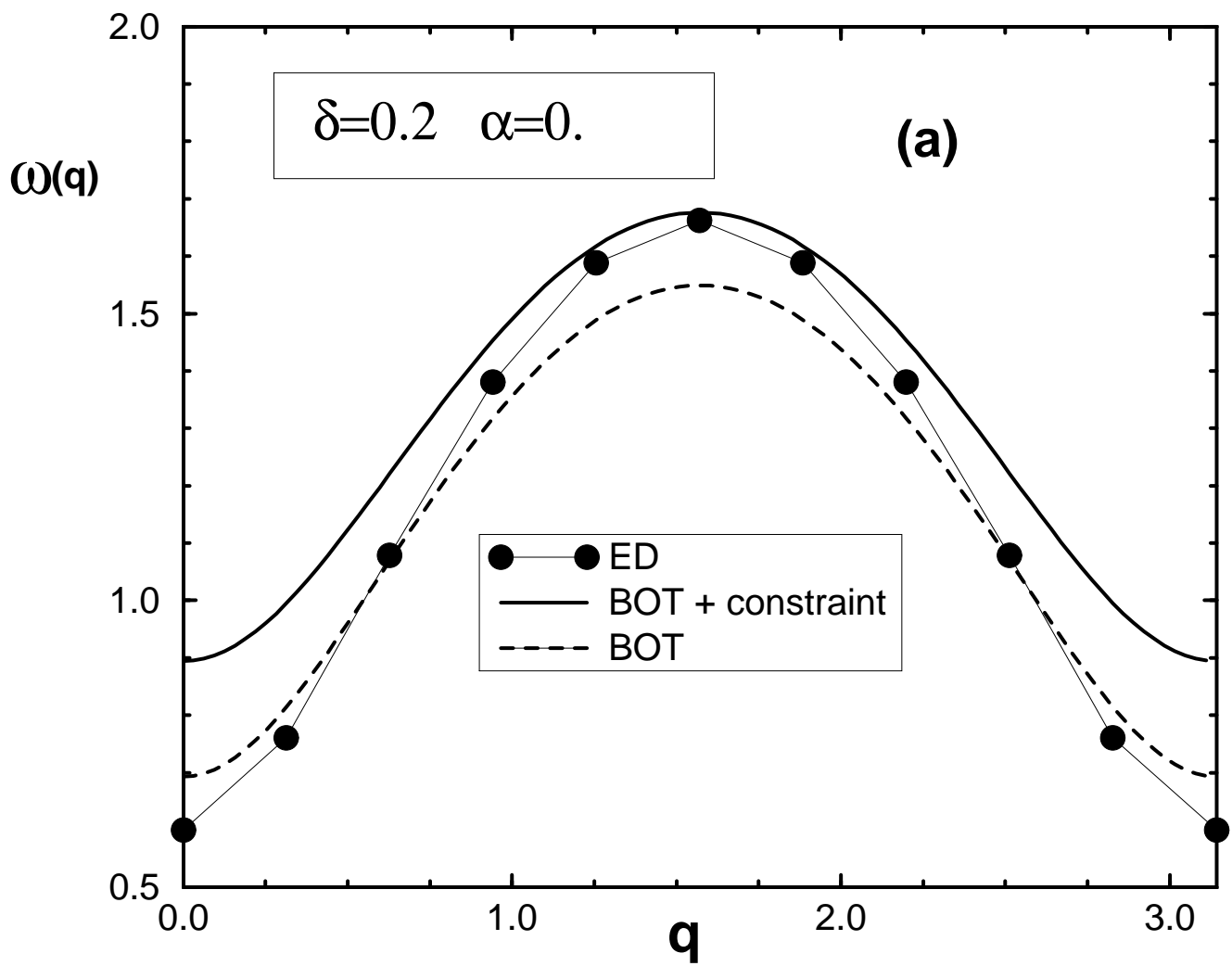
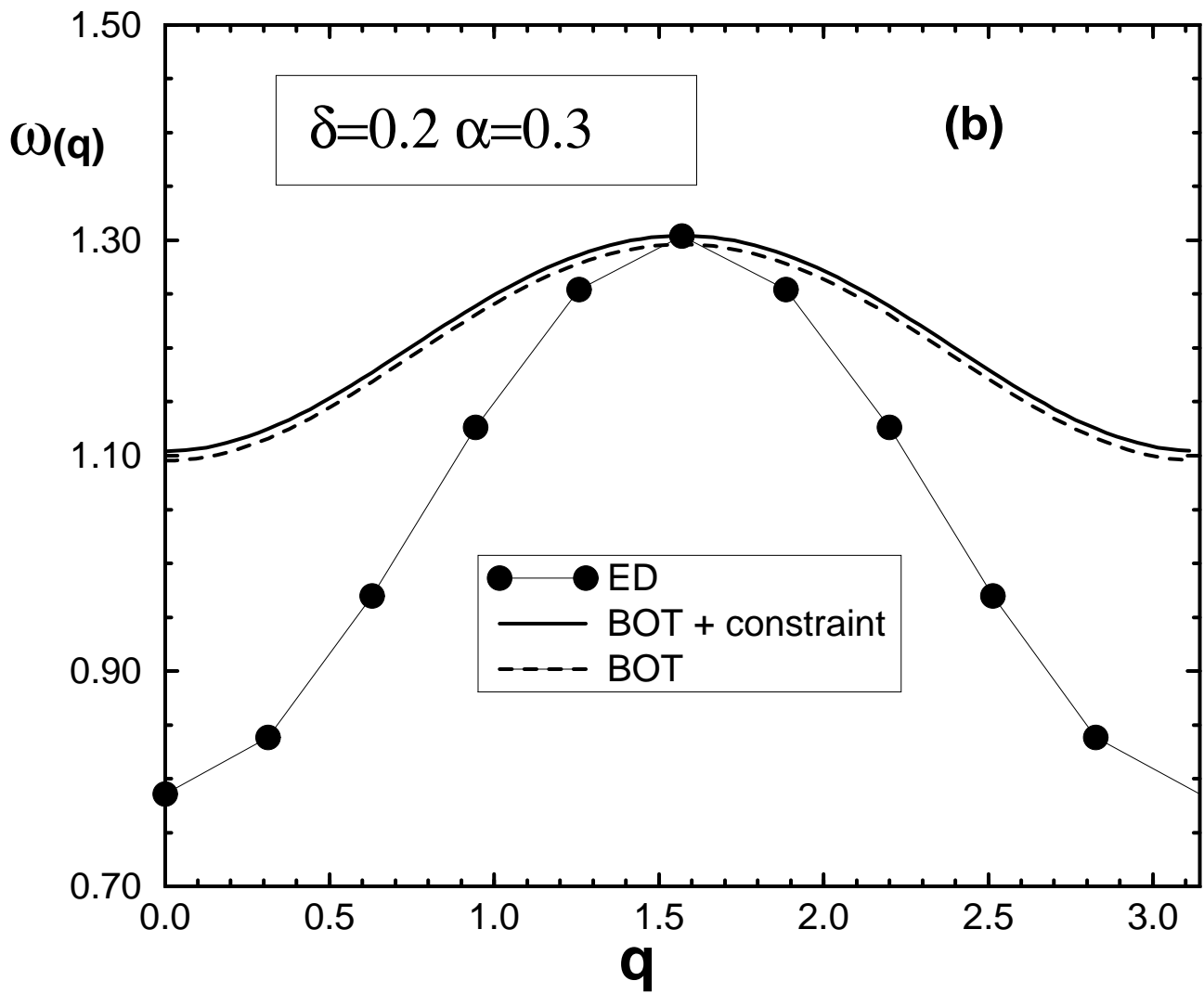
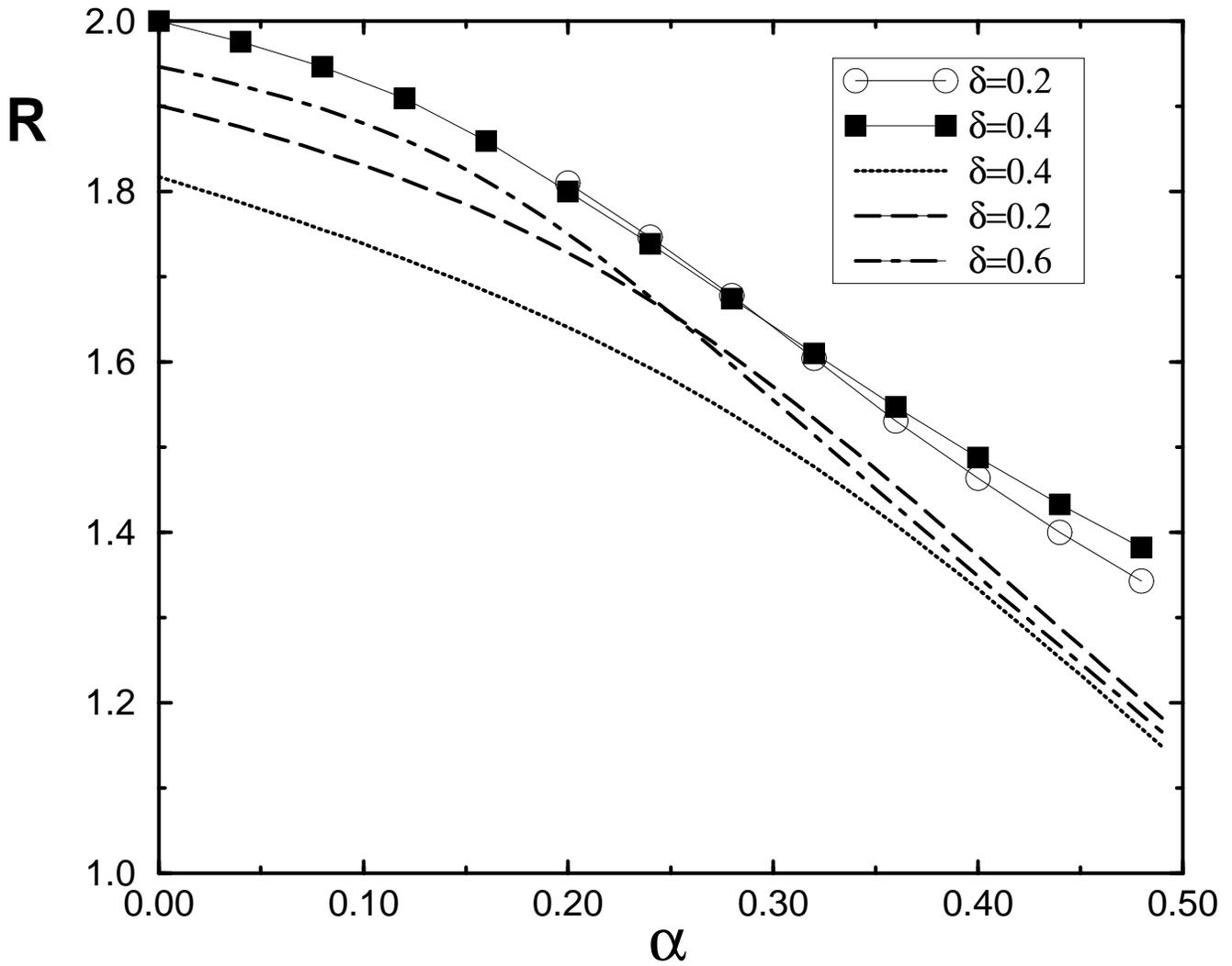
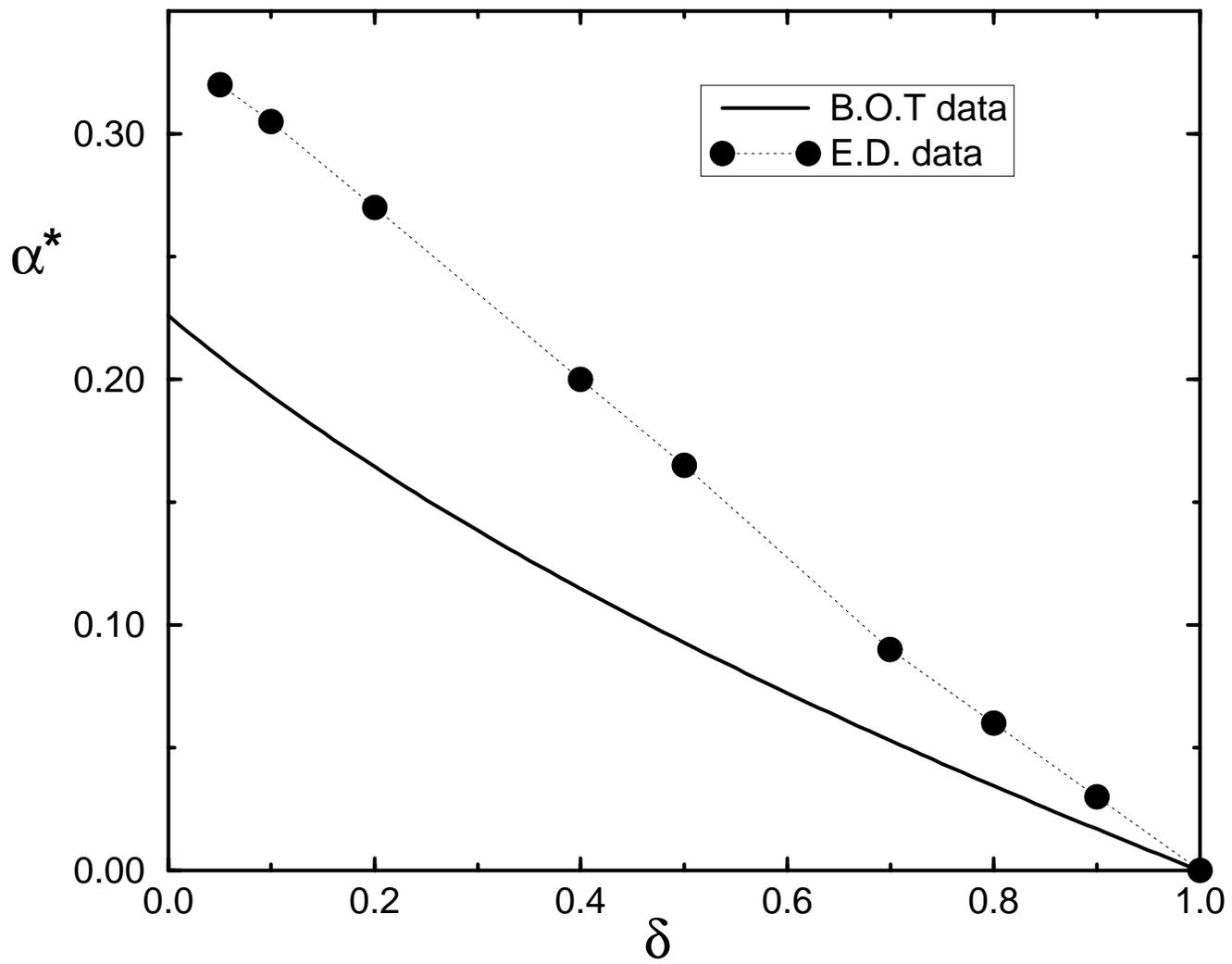
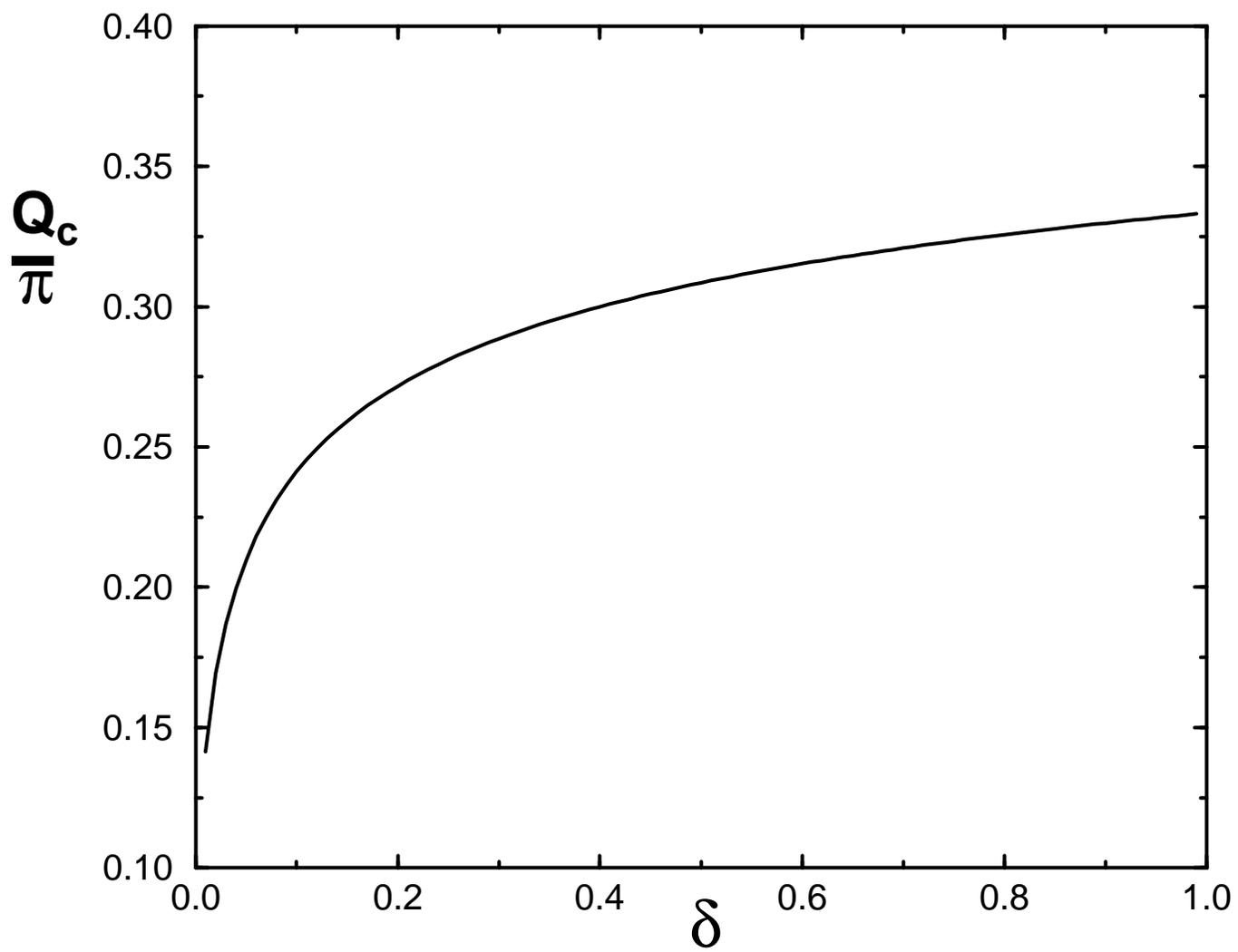
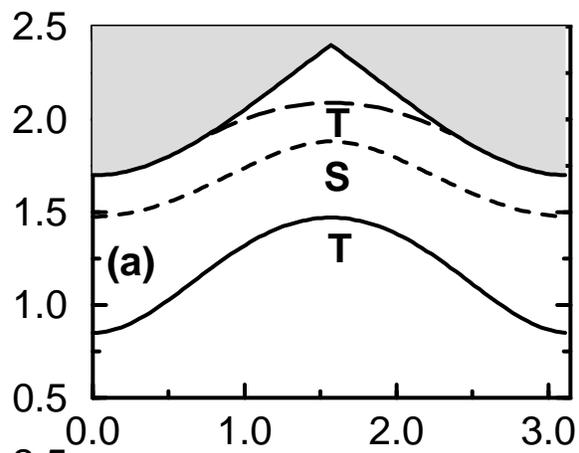
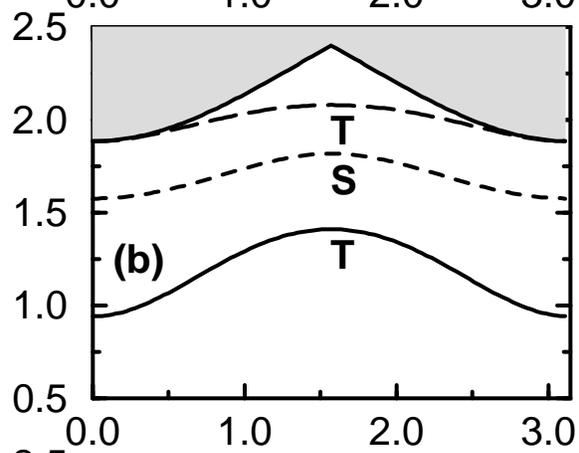
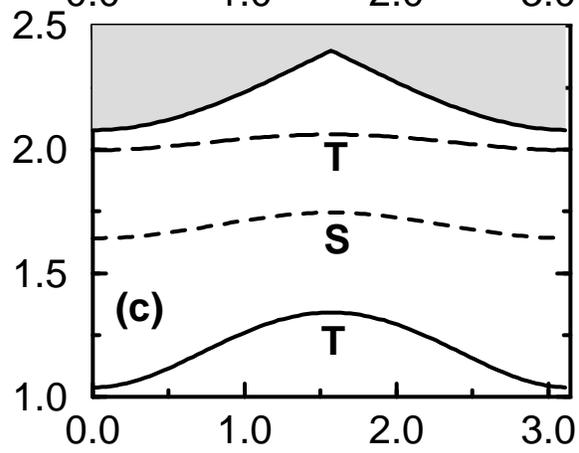
\end{document}